\documentstyle[12pt]{article}

\begin{document}
\begin{center}
{\Large \bf Ground state properties of dissipative two state
system coupled to phonons}\\
\vskip 0.8  true in

{\large T. P. Pareek$^1$ and A. M. Jayannavar$^2$}\\

{\small Institute of Physics, Bhubaneswar-751005, India}\\
\vspace*{0.6 in}
\today
\end{center}
\footnotetext[1]{e-mail: pareek@iopb.ernet.in}
\footnotetext[2]{e-mail: Jayan@iopb.ernet.in}

\begin{center}
ABSTRACT
\end{center}
We present a systematic analysis of ground state properties of a
tunneling system coupled to an Ohmic bath. The nonadiabatic
effects induced by coupling with tunneling particle , on the
ground state properties of phonons are taken into account. For
this we have considered both displacement and deformation of
phonons via a variational treatment. The symmetry breaking
transition , or the delocalization-localization  transition of
tunneling particle is presented and is compared with earlier
known results. We also show that in the critical regime a gap
will be opened near the zero momentum in the spectrum of the
phonons. The modified phonon spectrum along with the phonon
displacement field correlation functions have been calculated.
It is shown that in the critical regime the tunneling system
stabilizes the lattice against long wavelength fluctuations.

PACS NO: 73.40.Gk , 74.50.+r , 03.65 -$\omega$
\newpage
\begin{center}
\large {\bf 1. Introduction}
\end{center}
\par The statics and dynamics of two state tunneling system coupled
to an environment is of fundamental interest both in physics and
chemistry [1]. This rather simple system has been used to study the
motion of impurities (atoms or groups of atoms)in solids and
glasses which play an important role in determining low temperature
properties [2,3] , defect motion in alkali halides [4] ,
charged particle motion in metals, magnetic
impurities in metals [5,6] and molecular transition in liquids
[7] etc. In all examples mentioned above tunneling is a
many-body problem in the sense that the considered particle
couples to a large number of degrees of freedom arising from
medium or environment.In recent years there has been a growing intrest in the
influence of dissipation on the behaviour of macroscopic quantum
variables and quantum tunneling out of metastable state. In
particular,  the trapped magnetic flux threading a
superconducting ring interrupted by  weakly coupled
Josephson junction is expected to show macroscopic quantum
coherence. In this case the problem of tunneling of quantized
magnetic flux reduces to quantum motion of a particle in a
double well potential [1]. The existence of chiral molecules by a
superselection rule which origanates from the ever present coupling
of the molecule to the radiation field can be understood in the
frame work of two state system [8]. Here the two states are nothing
but the left and right handed configurations of molecule.
Recently the two state system formulation has been applied to an
understanding of c-axis resistivity in high-$T_c$ layered oxides
and transverse resistivity in quasi one-dimensional
compounds [9,10] . Electrical transport along c-axis in layered oxides is
pictured as a coherent interplanar tunneling between
neighbouring layers blocked by repeated intraplanar
incoherent(inelastic) scattering. Here two states correspond to
two nearest neighbour layers. The ground state properties of two
level system are also related to anisotropic Kondo problem [1] and
one dimensional inverse square Ising model [11,12] . In particular the
delocalization-localization transition in two state system is
related to the existence of ferromagnetic phase transition in
the one-dimensional inverse square Ising model and
several rigorous results for this problem are known.

\par In all these systems it is now well established that the
environmental coupling to tunneling system, makes a tunneling
motion less frequent and consequently reduces the effective
tunneling matrix element, sometimes termed as Debye-Waller
factor. This reduction stems from orthogonal properties of many
particle wavefunctions with different local potentials and is
known as Anderson orthogonality theorem [13] . In our study we examine
the ground state properties of two state system linearly coupled
to a phonon bath which provides Ohmic dissipation. This system
can be described by a spin-boson Hamiltonian.
\begin{equation}
\label{a}
H=-\Delta_o\sigma_x+\sum_k\omega_k a^\dagger_k a_k
+\sum_k g_k(a_k+a^\dagger_{-k})\sigma_z\quad ,
\end{equation}
where $\Delta_o$ represents bare tunneling matrix
element, $\sigma_x,\sigma_y$ are Pauli matrices. The bath is
described by a set of harmonic oscillators and with coupling
constant  $g_k$  and $a_k, a^\dagger_k$ are the annihilation
and creation operators representing phonon modes. We have taken
coupling constant $g_k$ of the form $g_k=g_o\mid{k}\mid^{1/2}$ ,
which represents coupling of a tunneling system to a
one-dimensional phonons. The frequency dispersion is taken to be
$\omega_k=\mid{k}\mid$ (acoustic phonons) , where k is the quasimomentum
wavevector of phonons. The Hamiltonian(\ref{a}) can also be
written in terms of creation ($c_i^\dagger$ )and annihilation
($c_i$) operators of the tunneling particle, namely
\begin{equation}
\label{b}
H=-\Delta_o(c^\dagger_1c_2+c^\dagger_2c_1)
+\sum_k\omega_ka^\dagger_ka_k
+\sum_kg_k(c^\dagger_1c_1-c^\dagger_2c_2)(a_k+a^\dagger_{-k})\quad,
\end{equation}
where 1 and 2 represent two equivalent positions of tunneling
particle. Since we are dealing with a single particle it does not
matter whether it is a boson or a fermion. The Hamiltonian for two
state system is also a good representation for statics and
dynamics of a quantum particle moving in symmetric double well
potential ,i.e, in the limit of low temperatures, $k_BT$  much smaller than
$\hbar\omega_o$ ($\omega_0$ the frequency of small amplitude oscillations in
the
wells ) and $V_o$ the barrier height between the wells [1]. The
tunneling frequency $\Delta_o$ is of the order of
$[\sqrt{\omega_oV_o/\hbar}] exp(-2V_o/\hbar\omega_o)$ , in the WKB limit.
The Ohmic dissipative bath corresponds to a special case [1]  where
the spectral density S($\omega$)is given by

\begin{equation}
\label{c}
S(\omega)=\sum_k g_k^2\delta(\omega-\omega_k)=
{\frac{1}{2}}\alpha\omega\quad,
\end{equation}
where  $\alpha$   is the dimensionless dissipation parameter.

\par The above Hamiltonian exhibits a rich phenomena for a tunneling
system ranging from damped oscillations(coherence) up to
localization [1] . It has been shown that for the weak coupling the
particle tunnels back and forth between two states and quickly
looses the memory of the initial quantum state. As the coupling
exceeds the critical strength , spontaneous symmetry breaking
occurs , leading to localization where the effective
tunneling matrix is identically zero [1,14,15] . This symmetry
breaking transition occurs only in the limit of zero
temperature.At finite temperature and for larger coupling
strength the tunneling particle hops incoherently from one state
to another. The zero temperature localization phenomenon
is associated with the infrared divergence induced by low energy
phonons ,i.e, the phonon field keeps the particle localized at
the expense of generating an infinite number of low-energy
phonons. The low freqency phonons play a dominant role in the
statics as well as in the dynamics of the two-state system. The
conventional weak coupling picture has a rather limited range of
validity and the adiabatic approximation over estimates the
occurence of infrared divergence.

Using the renormalization-group procedure the delocalization to
localization transition has been studied [1]. When the bare
tunneling matrix element $\Delta_o$ is small compared to the
upper phonon cutoff frequecy $\omega_c$  ,  the effective
tunneling matrix element is given by  $\Delta_{eff} =
\Delta_o({\frac{2e\Delta_o}{\omega_c})^{{\frac{\alpha}{(1-\alpha)}}}}
$ .
At $\alpha=1$ there is a transition from delocalized to
localized ($\Delta_{eff}=0 $) state. The renormalization-group results
have been obtained mainly based on the assumption of a dilute
instanton gas or dilute non-interacting blip approximation [1] . In
the later studies on dissipative quantum tunneling ,  it has been
shown that this approximation breaks down due to the condensation
of instanton gas [16] . It has also been pointed out that the
non-interacting blip approximation leads to a complete neglect of
reaction field which becomes important for the long time
dynamics in the larger coupling domain [17]. Silbey and Harris [18], and
Tanaka and Sakurai [19] have independently succeeded in rederiving
the renormalization-group results by a simple variational
procedure which amounts to a kind of mean field theory.  They have
taken the displacement of phonons in a ground state via
variational parameters.  It should be noted that when a particle
tunnels from one state to another it has two distinct effects on
the phonon wave function.  One is that of a displacement and the
other is that of a deformation.  High-frequency phonons ($\omega >
\Delta_o$) follow the tunneling particle adiabatically.  Hence
the displaced phonons gives the correct result in this domain.
However , for low-frequency modes and in the strong coupling regime such an
approximation is not sufficient.  The low frequency phonons in
general are not always in equilibrium with the particle motion ,
and hence the tunneling particle not only moves in an adiabatic
potential , but also gets influenced by a retarted potential
arising due to the phonon bath [4,21]. The motion of the particle at time
$t^\prime$ disturbs the phonon  state , which in turn acts on the motion of
the particle at later time  t. For the frequencies
$\omega < \Delta_o $ , these non-adiabatic effects dominate ,
which in turn apart from displacement produce strong deformation
in the ground state phonon wave function.  Chen and coworkers have treated
these effects via a variational treatment by  introducing a
squeezing parameter for the phonon states [20,22-24].  Within this treatment
they have analysed the symmetry breaking transition and have also
shown that the ground state properties of the tunneling particle
not only depends on the spectral density , but also on the
explicit form of the coupling strength on frequency of phonon
modes [20]. This non-universal behaviour is at variance with the
earlier results based on the renormalization-group and the
obtained phase diagram is in no simple way related to the known
results [1]. In Ref[25] , the phase diagram for the
delocalization-localization transition has been analysed through a
variational treatment which incorporates both the displacement
and the squeezing of wave function via two independent variational
parameters for each phonon mode.

\par Most of the theoretical treatments to date are concerned with
the effect of phonon bath on the tunneling motion. Very few
treatments  exist which directly address the question of effect of
tunneling particle on the phonon bath [20,26,27]. It has been shown that
[20] two state coupled system at zero temprature can be mapped on to the
modified quantum sine-Gordon model , which describes the effective non-linear
interaction between the low frequency phonons
when the tunneling system is in its ground state.
It is also of interest to know
that the ground state properties of ($ 1+1 $) dimensional
quantum sine-Gordon model can be mapped on to several two
dimensional statistical mechanics models ,e.g., two
dimensional Coulomb gas etc [27].  In the critical regime , where the
particle is still in the delocalized state , a gap in the
elementary excitation of the low frequency phonons is opened
near zero momentum , and the corresponding ground state wave
function is found to  be of a pairing quasi-particle state ,
analogous to the BCS super conducting state [26,27].

\par In our present work we have carried out a systematic analysis of
ground state properties of the tunneling particle as well as of the
phonons via a variational treatment which correctly treats both
the displacement and deformation of each phonon wave function in
its ground state . As stated earlier we have restricted our
treatment to a two state system , coupled to bosonic enviroment,
with a Ohmic spectral density.
Results do depend on specific form of dependence of
coupling  constant on the frequency of phonon mode, in our case
$ g_k = g_o \sqrt\omega_k $ , this coupling is common to
electron-phonon  coupling in solids. The reduction in the
effective tunneling matrix element of the particle as a function
of dimensionless dissipation parameter $\alpha $ is calculated
and is compared with earlier known results. The modified phase
diagram for the symmetry breaking transtion is presented. We have also shown
that in the critical regime and in the delocalized phase the gap
will be opened near zero momentum in the spectrum of the low
frequency phonons. The gap vanishes throughout in the localized
domain. An analytical expression for the phonon frequency gap as a function of
effective tunneling matrix element and dissipation parameter
$\alpha $ is obtained. We have also obtained an information
about the modified phonon spectrum and using this , the  phonon displacement
field spatial correlation function has been calculated. In the
delocalized phase the phonon field correlation functions at
large distances (in the asymptotic domain ) falls of
exponentially. This implies that in the presence of a tunneling
particle the  lattice is stable under the long wave length
fluctuations , so long as effective tunneling matrix element for
a particle is non-zero.

\begin{center}
\large {\bf 2. Ground state variational wave function}
\end{center}
In the following we first motivate our ground state variational
wave function of the system described by the Hamiltonian given
in eqn $(1)$ or $(2)$ . Ground state wave function for this
system can not be obtained analytically. Only special cases can
be solved exactly. The simple cases being $ (i) \Delta_o =0 $ ,
and $(ii)$ the coupling constant $ g_k=0 $ . In the first case
where $\Delta_o = 0 $ , the system is just an oscillator and
there is no tunneling motion between the two degenerate states.
However ,the oscillators are displaced in one direction when the
tunneling system is in one of the two levels  and are displaced
in the opposite direction when the tunneling system is in the other level.
Correspondingly the total system in its ground state can be
represented by a doubly degenerate localized state
\begin{equation}
\label{d}
\phi_i=\prod_kexp(-{\frac{1}{2}}{\frac{g^2_k}{\omega_k^2}})\quad
exp\{(-1)^i{\frac{g_k}
{\omega_k}}b^\dagger_k\}\mid0\rangle\quad.
\end{equation}
\noindent\\
Here $\mid 0 \rangle $ represents the vacuum state and i being
the position of the tunneling state. For an Ohmic dissipative
bath the overlap integral between the above two states tends to
zero , if there is no lower cutoff frequency , i.e, the two
ground states are orthogonal to each other. In the other cases  where $
g_k = 0$,i.e., tunneling particle and phonon modes decouple ,
the eigenstates of the tunneling particle are given by the
symmetric and antisymmetric combination of two level states ,
the symmetric combination being the ground state.  These two
situations indicate that for a finite value of $ g_k $ and
$\Delta_o $ the system exhibits the competition between the
localization resulting from the phonon modes and the
delocalization due to the tunneling. The particle tunnels from
one state to another , carrying a dressed phonon cloud. In view
of these facts , we immediately write down a ground state
variational wave function as
\begin{eqnarray}
\label{e}
\psi={\frac{1}{\sqrt2}}\left\{\prod_kexp[-{\frac{1}{2}}\mid{u_k}\mid^2]\quad
exp[-u_ka^\dagger_k]\quad{exp[-\gamma_k(a_ka_{-k}-a^\dagger_ka^\dagger_{-k})]}
C^\dagger_1\right .\nonumber\\
\left . +\prod_kexp[-{\frac{1}{2}}\mid{u_k}\mid^2]\quad
exp[u_ka^\dagger_k]\quad{exp[-\gamma_k(a_ka_{-k}
-a^\dagger_ka^\dagger_{-k})]}
C^\dagger_2\right\}\mid{0}\rangle\quad,
\end{eqnarray}
\noindent\\
where $u_k$ and $\gamma_k$ are both variational parameters for
each phonon mode.
We call our variational wave function (eqn.[\ref{e}]) as VWF(3) . When
the parameter $\gamma_k$ takes  a finite non zero value we say
that the corresponding phonon state (or two phonon state) is
squeezed [28]. The squeezed state represents a non-classical minimum
uncertanity state and are eigenstates of a linear combination of
the annihilation and creation operators. It differs from a minimum uncertainty
coherent state in that it has uncertainty less than that of a coherent
state in one of the quadrature phase given by $exp(-\gamma_k) $
and in another $exp(\gamma_k) $ , which means that one of the
quadrature phase is squeezed at the expense of other. In real
space representation the displaced squeezed state is given as
$\phi(x)=[\pi S^2]^{-1/4}\quad exp[-(x-x_0)^2/2S^2]$, where $x_0$
is the mean displacement and $S=exp(\gamma_k)$. For this state
the uncertainty , $\Delta{x}$ in spatial coordinate equals to
$exp(\gamma_k)$ and $\Delta{p}$ in the momentum equals
$exp(-\gamma_k)$. If $\gamma_k$ takes a positive value the wave
function spreads over a larger distance than the ground state
wave function of the harmonic oscillator (coherent state).
Squeezing parameter takes care of the deformation associated
with the low frequency phonon wave function as discussed
earlier. In VWF(3) if we substitute $\gamma_k =0$ , we get the
variational wave function proposed independently by Tanaka and
Sakurai [19] and Silbey and Harris [18] , we call this VWF(1).
In the VWF(1) , if we substitue $u_k=0$ , it becomes an exact
ground state of the unperturbed tunneling particle. On the other
hand if we put $u_k= -g_k/w_k $, the wave function describes a
linear combination of two degenerate ground states given in eqn(\ref{d}).
The VWF(1) does not take into account the deformation of low frequency phonons.
In VWF(3) if  we take varitional parameter $u_k$ to be constant and equal to
$(-g_k/\omega_k)$ , we get the variational wave function
proposed by Chen et.al [20,22-24] . We call this  as a VWF(2).
The VWF(2) takes in to account the deformation of low frequency
phonon via squeezing parameter $\gamma_k$ . However , it does
not take properly in to account the displacement of phonons. All
the phonons are assumed to be displaced by an amount $\mid{u_k}\mid =
g_k/\omega_k$ , which is true only in the adiabatic limit or for
phonon frequency $\omega > \Delta_o $. In the VWF(3) both $u_k$
and $\gamma_k$ being the variational parameter , naturally it
takes care of non adiabatic effects in both the displacement and
the deformation of low frequency phonons properly.

\par Now using VWF(3) one can obtain the expectation value of
the total ground state energy ($E_g$) . Using the well known
properties [20,22-30] of squeezed states and bose operators we get,
\begin{eqnarray}
\label{f}
E_g(u_k , \gamma_k) = -\Delta_o\left\{exp\sum_k[(-2u_k^2)
exp(-4\gamma_k)]\right\}\nonumber\\
+\sum_k\omega_k[u_k^2+(sinh2\gamma_k)^2]+\sum_k2g_ku_k\quad.
\end{eqnarray}
The variational parameters $u_k $ and $ \gamma_k $ are
determined by minimising $ E_g $ with respect to $ u_k $ and $
\gamma_k $ , i.e. , we set ${\frac{\partial{E_g}}{\partial{u_k}}} = 0$ and
 ${\frac{\partial{E_g}}{\partial{\gamma_k}}} = 0$. From this we obtain
\begin{equation}
\label{g}
u_k=\left[-g_k/(2\Delta_0Zexp(-4\gamma_k)+\omega_k)\right]\quad,
\end{equation}
\begin{equation}
\label{h}
\gamma_k=(1/8)ln\left[1+8\Delta_oZu^2_k/\omega_k\right]\quad.
\end{equation}
\noindent The factor multiplying $\Delta_o$ in the eqn. (6) is
identified as a tunneling reduction factor Z (or Debye-Waller
factor ), and is given by
\begin{equation}
\label{i}
Z=exp\left\{\sum_k[-2u_k^2exp(-4\gamma_k)]\right\}\quad.
\end{equation}
\noindent The effective matrix element $\Delta_{eff} $ is given
by $\Delta_{eff}=\Delta_oZ$ . The reduction factor Z can be
determined by using eqns.(\ref{h}) and (\ref{i})
self-consistently. One can notice from eqns.(\ref{g}) and
(\ref{h}) that $u_k$ and $\gamma_k$ tends to $(-g_k/\omega_k)$
and zero respectively in the high frequency limit as discussed
earlier. This is consistent with an adiabatic treatment in this
frequency domain , i.e. , high frequency phonons with $\omega$
greater than the $\Delta_{eff}$ follow the tunneling particle
adiabatically.

\begin{center}
\large {\bf 3. Effective phonon Hamiltonian}
\end{center}

\par We will now derive an effective Hamiltonian for phonon bath at
zero temperature consistent with the VWF(3). To this end we first
perform a canonical transformation

\begin{equation}
\label{j}
U_1=exp\left[\sigma_z\sum_ku_k(a_k-a^\dagger_{-k})\right]\quad,
\end{equation}
\noindent on Hamiltonian (\ref{a}). we get
\begin{eqnarray}
\label{h1}
H_1=U^\dagger_1HU_1=
\sum_k\omega_ka^\dagger_ka_k
-\Delta_o\sigma_xCosh\left[\sum_k2u_k(a_k-a^\dagger_{-k})\right]\nonumber\\
+i\Delta_o\sigma_ySinh\left[\sum_k2u_k(a_k-a^\dagger_{-k})\right]\nonumber\\
+\sum_k(g_k+u_k\omega_k)\sigma_z+2\sum_k(u^2_k\omega_k+g_ku_k)\quad.
\end{eqnarray}
{}From the Hamiltonian in eqn.(\ref{h1}) we see that there are two
distinct types of influence on the phonon bath due to coupling
with two-state system. The first is the diagonal term contaning
$\sigma_x$ , giving static influence due to two level system being its
ground state $(\sigma_x=+1)$. The nondiagonal interaction terms
contaning $\sigma_y$ and $\sigma_z$ represents the dynamic
influence due to the transition in two state system  between its
ground state and excited state on the phonon system.
These nondiagonal terms are
important  only at finite temperatures. Hence at zero
temperature we neglect these terms along with the constant
factor appearing in eqn.(\ref{h1}) , thus we have
\begin{equation}
\label{h2}
H_{eff}=\sum_k\omega_ka^\dagger_ka_k
-\Delta_ocosh\left[\sum_k2u_k(a_k-a^\dagger_{-k})\right]\quad.
\end{equation}
\noindent Equation.(\ref{h2}) represents an effective phonon
Hamiltonian , having a nonlinear interactions between the
phonons. Here nonlinear interactions are not only between the
phonon in the same mode , but also in the different modes.
By appropriately defining the field operators one can readily
map the $H_{eff}$ on to modified quantum sine-Gordon model (see
ref[20]) in 1+1 space-time dimension. It is well known that the
quantum sine-Gordon model exihibits a Coleman phase trasition as
a function of the coupling constant. In Ref[26,27] critical
conditions for this phase trasition have been derived. It is
also shown that , in the critical regime the gap will be opened
in the excitation spectrum near zero momentum . The Ref[27],
also elaborates on the connection with Kosterlitz-Thouless classical
theory of two dimensional Couloumb gas. In these references
[20,26,27] static properties of two state systems have been
studied , by mapping to sine-Gordon model. The results obtained
are in spirit of VWF(2). However , our results differ in that we
closely follow the spirit of VWF(3) , which itself is a better
ground state variational wave function.
For the effective phonon system (eqn.[\ref{h2}]) , we apply the second
unitary transformation
\begin{equation}
\label{u2}
U_2=exp\left[\sum_k\gamma_k(a_ka_{-k}-a^\dagger_ka^\dagger_{-k})\right]\quad.
\end{equation}
\noindent  This transforms the effective Hamiltonian (\ref{h2}) into
\begin{eqnarray}
\label{h3}
H_{eff}=\sum_k\omega_k[b^\dagger_kb_kcosh4\gamma_k
+(sinh2\gamma_k)^2 \nonumber\\
+(1/2)(b^\dagger_kb^\dagger_{-k}+b_kb_{-k})(sinh4\gamma_k)]\nonumber\\
-\Delta_ocosh[\sum_k2u_k(b_k-b^\dagger_{-k})
exp(-2\gamma_k)]\quad,
\end{eqnarray}
\noindent where $b_k$ and $b^\dagger_k$ are
\begin{equation}
\label{b1}
b_k = U^\dagger_2a_kU_2 = a_k cosh2\gamma_k -
a^\dagger_{-k}sinh2\gamma_k\quad,
\end{equation}
\begin{equation}
\label{b2}
b^\dagger_k = U^\dagger_2a^\dagger_kU_2 =
a^\dagger_kcosh2\gamma_k-a_{-k}sinh2\gamma_k\quad.
\end{equation}
The transformation (\ref{u2}) some times refered to as a
Bogolibov transformation, basically it resacles the phonon
normal-mode coordinate and momenta.
We can expand the hyperbolic function in eqn.(\ref{h3}) by normal
ordering phonon operators. We choose $\gamma_k$ such that it
minimises the ground state energy
as given by eqn.(\ref{h}). With
this choice of $\gamma_k$ we get , after the
straight-forward algebra (see also ref.[27])
\begin{equation}
\label{h7}
H_{eff}=\sum_k\omega_kexp(4\gamma_k)b^\dagger_kb_k+E_o\nonumber\\
-\Delta_oZ[(i\phi_+)^4/240-(i\phi_+)^3(i\phi_-)/6+....]\quad,
\end{equation}

\noindent where $\phi_+$ and $\phi_-$ are
\begin{equation}
\label{h5}
\phi_+=\sum_ku_kexp(2\gamma_k)b^\dagger_{-k}\quad,
\end{equation}
\begin{equation}
\label{h6}
\phi_-=\sum_ku_kexp(-2\gamma_k)b_k\quad.
\end{equation}

\noindent  The ground states of  (\ref{h7})  are displaced squeezed
states in the original phonon bases.
\begin{equation}
\label{h8}
\phi=exp\{\sum_k\gamma_k(a_ka_{-k}-a^\dagger_ka^\dagger_{-k})\}\quad
exp\{-\sigma_z\sum_ku_k(a^\dagger_{-k}-a_k)\}\mid0\rangle\quad.
\end{equation}

We see that Hamiltonian (\ref{h7}) corresponds to a free field
model if the tunneling reduction factor Z approaches zero. The
effective tunneling matrix element approaching zero implies the
symmetry breaking transition for a tunneling particle in that
the particle goes from the delocalized state to a localized
state .The effective tunneling matrix element
$\Delta_{eff}\rightarrow 0$ also implies the vanishing of
nonlinear interaction terms in Hamiltonian (\ref{h7}) as well
as in the Hamiltonian (\ref{h2}) . In eqn. (\ref{h7}) we
identify $\Omega(k) = \omega_kexp(4\gamma_k) $ as the new excitation
spectrum of the phonon quasiparticles with their ground state wave function
given by eqn.(\ref{h8}) . It is clear that in the critical regime where Z is
small $(Z << 1)$ , the higher order terms in eqn.(\ref{h7}) are small and
correspondingly $\Omega(k) $ is well defined excitation spectrum
of a modified phonon bath in the presence of a tunneling particle. In this
limit we later show that the gap will be opened in the vicinity
of the zero momentum in the excitation spectrum of the phonons
and this gap will disappear as $ Z\rightarrow 0$ . The opening up of the gap in
the
vicinity of the zero momentum naturally takes care of infrared
divergence problem arising in a effective tunneling
matrix element without invoking infrared cutoff prescription.
With a modified excitation spectrum near the critical regime ,
we will also calculate the phonon displacement field correlation functions.

\begin{center}
\large {\bf 4. Numerical procedure }
\end{center}
\par We will first outline the numerical technique used to
investigate the behaviour of the reduction factor from the
expression for Z (eqn[\ref{i}]) . For the case of Ohmic
dissipation this can be written as
\begin{equation}
\label{h10}
lnZ=-\alpha\int_o^{\omega_c}d\omega\omega exp(-4\gamma_k)/
(2\Delta_oZexp(-4\gamma_k)+\omega)^2\quad,
\end{equation}
\noindent where $\alpha = g_o^2/\pi $ is a dimensionless
dissipation parameter, which characterises the coupling between
the particle and the surrounding medium and $\omega_c$ is the upper phonon
frequency cut-off. Throughout we have taken $\omega_k=\mid k \mid$.
To evaluate Z we have to use $u_k$ and $\gamma_k$
as given in eqns.(\ref{g}) and (\ref{h}) respectively. The expression for
$\gamma_k$ contains $u_k$ and vice-versa. Substituting these
values for $u_k$ and $\gamma_k$ in
eqn.(\ref{h10}) leads to an infinite (continued fraction) expression. In
ref [25] , this self-consistent equation has been solved by
iterative procedure where one terminates the infinite heirarchy
to a finite one. This procedure is a approximate one.
We have developed an exact numerical procedure
without resorting to any approximations. To this end we define a
new quantity $\Omega(k) = \omega_kexp(4\gamma_k) $ with this definition
eqn.(\ref{h10})  reduces to
\begin{equation}
\label{o1}
lnZ=-\alpha\int_o^1d\omega \Omega(k)/(2\Delta_oZ+\Omega(k))^2\quad.
\end{equation}
\noindent Using eqns (7) and (8) one can readily show that $\Omega(k)$ obeys a
forth order polynomial equation , namely
\newpage
\begin{eqnarray}
\label{o2}
\Omega^4+4\Delta_0Z\Omega^3
+(4\Delta_0^2Z^2-8\Delta_0Zg_k^2/\omega_k-\omega_k^2)\Omega^2\nonumber\\
(4\omega_k^2\Delta_0Z)\Omega-4\omega_k^2\Delta_0^2Z^2=0\quad.
\end{eqnarray}
\noindent Here $g_k=g_o\mid \omega_k \mid^{1/2}$.
In eqns.(\ref{o1}) and (\ref{o2}) we have rescaled the
energy with respect to $\omega_c$, for example $\Delta_o =
\Delta_o/\omega_c$ , $\Omega(k)=\Omega(k)/\omega_c $ etc.
For a given value of $\alpha$($=g_k^2/\pi$) , and Z this equation
gives four solutions. It turns out that out of these four
solutions two are complex and one is positive and another is
negative. We take only the physical solution,i.e., positive
one as it represents the energy of the modified phonon modes.
In some parameter range where $\Delta_o/\omega_c$ is large, four
solution to $\Omega$ of eqn.(\ref{o2}) are all real. It turns
out that three of them are negative ,hence we again take a
positive solution.
We first take a fixed value of $\alpha$ , and $\Delta_o$ and
for a value of Z lying in an interval $1 < Z < 0 $ we
calculate $\Omega(k)$ for all $\omega_k$ . Substituting this $\Omega(k)$
for all $\omega_k$ in the expression (\ref{o1}) we check for
self-consistency. We start initially with Z=1 and go on changing
Z in a small step till we reach the self-consistent solution.

\newpage
\begin{center}
\large {\bf 5. Results and discussion }
\end{center}

\par In fig.(1) we have plotted the reduction factor Z as
a function of dimensionless parameter $\alpha$ for a fixed value
of $\Delta_o/\omega_c = 0.1 $. For comparison we have also
plotted the earlier results obtained from VWF(1) (....) and
VWF(2) (- - - ). In all these cases the reduction factor
decreases as expected as a function of $\alpha$ and at a
particular value $\alpha_c$ there is a symmetry breaking
transition from delocalized state to localized state, at which Z
becomes identically zero. For $\Delta_0/\omega_c = 0.1 $ the
VWF(1) gives $\alpha_c$ around 1, whereas VWF(2) gives $\alpha_c$
around 2. In our case we obtain $\alpha_c = 1.85 $. We have
identified $\alpha_c$ such that at this value Z becomes less
than $10^{-7}$. In parameter range of
$\alpha$ between zero and 0.5 the results obtained from VWF(1) and
VWF(3) are almost identical. As we approach $\alpha_c$ the results
deviate. This deviation should be identified as a result of inclusion
of fluctuations over the mean field results obtained earlier [1]. It
is also of interest to note that the parameter range $0 < \alpha <0.5$
in the limit $\Delta_o/\omega_c << 1$ is of most interest for the
quantum coherence problem [14]. In this regime at temperature T=0,
the tunneling system exhibits damped oscillations (coherence) with
frequency $\Delta_{eff}$. The symmetry breaking transition
obtained from VWF(3) is continous in the sense that the
reduction factor Z apporaches zero continously as $\alpha$ tends
to the critical value $\alpha_c$. Even for larger values of
$\Delta_o/\omega_c$ we have observed that the results obtained
for the reduction factor Z from VWF(1) and VWF(3) are identical
in the range of small dissipation parameter $\alpha <<1$. In the
same parameter range of $\alpha$ the results obtained from
VWF(2) and VWF(3) differ considerably. Moreover, it should be
noted that the results for Z obtained from VWF(1) in the limit
$\alpha<<1$ are cosistent with the known perturbation results [17].

\par In fig(2) we have plotted the phase boundary between the delocalized
(Z$\neq$0) and localized (Z = 0) phase for various values  of
$\Delta_o/\omega_c$. In the limit of
$\Delta_o/\omega_c\rightarrow 0$ the
phase boundary is located at $\alpha_c=1$. The result based on
instanton calculation or path integral calculation also predict
the phase boundary to be at $\alpha_c=1$ as $\Delta_o/\omega_c\rightarrow 0$.
It also predicts the phase boundary to be at
$\alpha_c=1$ till $\Delta_o/\omega_c$ reaches a value of $1/2e=
0.184$. In our case the phase boundary shifts to the
right side (increase in $\alpha_c $) continuously as we vary
$\Delta_o/\omega_c$. The VWF(2) predicts the phase boundary
starting from $\alpha_c=2$ in the limit
$\Delta_0/\omega_c\rightarrow 0$
to a higher values of $\alpha_c$ as $\Delta_o/\omega_c$
increases. In a sense the phase boundary  of the delocalization
to localization transition obtained from VWF(1) and VWF(2) are
not related. We can say with a certan confidence that our results
clearly shows the improvement over the earlier results obtained
by various treatment using path integral formulation [1]. In effect
our results amount to taking care of flucutations (via a
squeezing parameter) in an otherwise mean field results obtained
earlier.

\par From eqn.(\ref{h7}) we have identified $\Omega(k) =
\mid{k}\mid{exp(4\gamma_k)} $ as excitation spectrum of the modified phonon
bath under the coupling with a tunneling particle. This
excitation spectrum aquires a well defined meaning only when the
$\Delta_{eff} $ is small , in this regime the higher order
correction terms in eqn.(\ref{h7}) becomes quite small and give a
small correction to the excitation spectrum via perturbation. In
fig.(3) we have plotted the excitation spectrum $\Omega(k)$
versus wave vector k , for $\Delta_o/\omega_c=0.1$ and $\alpha = (.9599)$
for which Z=.1005 . We have calculated the excitation spectrum
with the help of eqn.(\ref{o2}) for a given value of $\alpha$ and
corresponding value of Z  which has been calculated by self
consistent procedure as mentioned earlier. The eqn.(\ref{o2})
for $\Omega(k)$  gives
four solutions and out of these there is only one physically
acceptable solution which gives a positive value of $\Omega(k)$.
We see from the fig.(3) that the gap opening
in the vicinity of zero momentum.
For comparision we have also plotted the excitation spectrum
$\omega_k=\mid{k}\mid$ of the unperturbed free phonon modes (in
the absence of coupling to a tunneling system). It is clear from
the figure that the low frequency phonon modes are affected due
to coupling as compared to high frequency modes as expected.
In fig.(4) we have plotted the
zero momentum excitation gap $\Omega(0)$ versus the
dimensionless dissipative cinstant $\alpha$ , for a fixed value of
$\Delta_o/\omega_c =0.1$. The gap has initially a value zero up to some
$\alpha$ , then it increases and after exhibiting a maxima the
gap vanishes identically at the critical value of $\alpha_c$.
In the regime where the gap $\Omega(0)$ is finite , it is given
by $\Omega(0) = -2\Delta_oZ+2\sqrt(2\Delta_oZ\alpha\pi)$ . Even
though in fig.(4) we have plotted $\Omega(0)$ versus $\alpha$
for the full range of $\alpha$ where tunneling system undergoes
delocalization to localization transition , the gap $\Omega(0)$
is meaningful quantity only in the critical regime , i.e ,
$\alpha$ close to $\alpha_c (Z << 1)$. If one would have carried
out the calculations in the spirit of VWF(2), i.e, by setting
the parameter $u_k=(-g_k/\omega_k)$ in the canonical
transformation given in eqn.(10), we can readily obtain [26,27]
an analytical expression for the excitation spectrum of the
modified phonon bath. This is given by
$\Omega(k)=\mid{k}\mid[1+(4\pi\alpha\Delta_oZ/k^2)]^{1/2}$ , with
a gap in the vicinity of zero momentum
$\Omega(0)=\sqrt4\pi\alpha\Delta_oZ$. The expression for
$\Omega(k)$ depends explicitly on $\alpha,Z,\Delta_o,and\quad k$. As
the results derived from VWF(2) and VWF(3) for Z and
consequently the symmmetry breaking transition  as a function of
$\alpha$ differ qualitatively, we have not made the comparision
between $\Omega(k)$ obtained in the spirit of VWF(2) and VWF(3).

\par Once we have the full information about the  spectrum one
can easily compute the phonon displacement field spatial
correlation function C(x). This correlation function is defined as [27]

\begin{equation}
C(x)={\frac{1}{2}}\int_o^{k_c}{\frac{dk}{2\pi}}{\frac{cos(k x)}{\Omega(k)}}.
\end{equation}
\noindent where $k_c$ is the upper phonon frequency cutoff $k_c=\omega_c$
In fig.(5) we have plotted C(x) versus x (=$xk_c$) for $\Delta_o/\omega_c =
0.1,
\alpha = (.9599) $ and Z = (.1005). The correlation function
intially follow a logrithmic behaviour
between the two space point with the increasing of their space
distance x.At the large distances
(or asymptotically) C(x) decays exponentially, after
showing a oscillatory behaviour in the intermediate regime.
The charcteristic decay length $\xi$ (or the correlation length)
is inversely proportional to $\Omega(0)$ and hence $\xi$ becomes
large in the critical regime and diverges at $\alpha_c$, the
transition point of delocalization to localization transition.
We have also observed on the expanded scale that the domain of
$x <<\xi$, the correlation function decays logarithmically.
It is important to note that in the absence of coupling of
tunneling particle to  lattice phonons , the phonon field
correlation function diverges logarithmically. This indicates the
instability of lattice against long wave length fluctuations.
However ,by coupling lattice to a tunneling system, the
lattice acquires stability against the long wave length
fluctuations so long as tunneling system is  in a delocalized
state ( i.e , when the excitation spectrum has a gap in the
vicinity of a zero momentum ). This fact manifests itself in
exponential decay of C(x) in the asymptotic regime . Thus a
single impurity ( or a tunneling system ) can stablize a one
dimensional lattice. The correlation length $\xi$ scales as $\xi\sim
Z^{-1/2}$ and it diverges as we approach $\alpha_c$ from below.

In conclusion we have studied the ground state properties of a
coupled spin-boson system i.e., a tunneling system coupled to an
Ohmic dissipative interaction arising due to a linear
interaction of tunneling system to a one-dimensional set of
harmonic phonons with an appropriate spectral density.
We have presented a modified phase
diagram for symmetry breaking transition. The results for the
behaviour of reduction factor on the dissipation parameter are
given and compared with known results. We have also shown
that the effective phonon system shows a gap in the excitation
spectrum in the critical regime and phonon displacement field correlation
function decays exponentially in the same regime , this
stabilizes lattice against the long wave length fluctuations.
Throughout our analysis we have restricted to a special form of
coupling constant , namely $ g_k$ is proportional to $\mid{k}\mid^{1/2}$. It is
known that the different forms for coupling constant
lead to qualitatively different results for symmetry breaking
trasition , in a sense there is no universality in a dissipative
two state system [21]. The zero temperature properties depend not
only on the phonon spectral density , but also on the explicit
dependence of coupling strength on the phonon frequency. We are
at present exploring the behaviour of effective phonon spectrum
for different dependence of coupling constant and the result will be
presented elsewhere.

\newpage
\par FIGURE CAPTIONS

Fig.(1).  Plot of reduction factor versus dimensionless
dissipation constant $\alpha$ , for a fixed value
$\Delta_o/\omega_c=0.1$. The ..... , - - - - , and ------ lines
are obtained from VWF(1), VWF(2) and VWF(3) respectively.

Fig.(2). Phase diagram for the delocalization-localization in
$(\Delta_o/\omega_c) - \alpha$ plane.

Fig.(3). The elementary excitation spectrum
$(\Omega(k)/\omega_c)$ of the low frequency phonons versus scaled
wave vector $k/k_c$ for a fixed value of $\Delta_o=0.1$,
$\alpha=.9599$ and the corresponding value of Z=0.1005. The
dotted lines(- - -) represent excitation spectrum
$\omega_k=\mid{k}\mid$, for uncoupled phonon modes.

Fig.(4). The zero wave vector gap $\Omega(0)/\omega_c$ in the
excitation spectrum versus dimensionless dissipation constant
$\alpha$ . For a fixed value of $\Delta_o=0.1$.

Fig.(5). The phonon field displacement spatial correlation
function C(x) versus scaled distance $k_cx$ for a fixed value of
$\Delta_o=0.1$, $\alpha=.9599$ and the corresponding Z=0.1005.

\end{document}